\documentstyle[psfig]{l-aa}

\newcommand{\be}{\begin{equation}}
\newcommand{\ee}{\end{equation}}
\newcommand{\bea}{\begin{eqnarray}}
\newcommand{\eea}{\end{eqnarray}}
\newcommand{\bfi}{\begin{figure}}
\newcommand{\efi}{\end{figure}}
\def\frac#1#2{{{#1}\over{#2}}}

\def\theta{\vartheta}
\def\phi{\varphi}
\def\epsilon{\varepsilon}

\def\ltsima{$\; \buildrel < \over \sim \;$}
\def\lsim{\lower.5ex\hbox{\ltsima}}
\def\gtsima{$\; \buildrel > \over \sim \;$}
\def\gsim{\lower.5ex\hbox{\gtsima}}

\begin{document}

\title{On the location of X--ray peaks and dominant galaxies in 
clusters}

\author{Davide Lazzati\inst{1,2} \and Guido Chincarini\inst{1,2}}

\offprints{Davide Lazzati}

\institute{
Osservatorio Astronomico di Brera, via E. Bianchi 46, I-23807 Merate (Lc)
Italy; 
\and Universit\`a degli Studi di Milano, via Celoria 16, I-20133 Milano Italy
\\ {\it e-mail:} \ lazzati, guido@merate.mi.astro.it
}

\date{Received ... ; Accepted ...}

\maketitle

\begin{abstract}

We analyze the structure of the X--ray emission 
of a sample of 22 Abell clusters of galaxies with a cD in their centre,
observed with the ROSAT PSPC. Utilizing the multi-scale
power of the Wavelet Transform we 
detect significant ($\sim 50 \, h^{-1}$ kpc) offsets 
between the large scale centroid and the peak of X--ray 
emission.
Despite the uncertainties on the satellite pointing, the X--ray to optical
correlation indicates a likely association
between the X--ray peak and the dominant galaxy.
We develop a model in which the offset is produced by small amplitude 
oscillations of the cD galaxy around the bottom of the cluster 
potential well and successfully 
compare it to the observed distribution of offsets.
Within this scenario the offsets are not due to 
dynamic instabilities and the number of structured clusters is greatly
reduced.

\end{abstract}


\section{Introduction}
\label{intro}

It is well known that the presence of a dominant galaxy (cD) in the
core of a galaxy cluster is a sign of dynamical evolution, being
highly correlated with a smooth spatial distribution
and a low spiral fraction of member galaxies (Sarazin 1988). 

In a large fraction of clusters with cDs, the X--ray peak is offset
with respect to the centroid of large scale emission
(Mohr et al. 1993, 1995).
Whether these offsets are caused by substructure, due to the
infall of smaller groups of galaxies, or not, is still an open issue. Indeed,
the fraction of clusters with substructure is strongly linked to 
the density parameter $\Omega_0$ (Richstone et al. 1992; 
Nakamura et al. 1995) and any spurious substructure would confuse
the potential link between observed cluster morphologies and that 
expected theoretically.

In a sample of 22 X--ray images of Abell clusters 
with cD galaxies in their centre (see section~\ref{data} for 
the description of the sample), we find that the peak of the X--ray 
emission is always located about the position of the dominant galaxy
and is offset from the large scale centroid.
In this work we propose an alternative to 
infall or Intra Cluster Medium (hereafter ICM) instability 
models to explain the presence of these offsets since:
a) these phenomena are unexpected in clusters which have the 
   appearance of being relaxed systems (cooling flows and dominant
   galaxies);
b) infall would disrupt the cooling flow and produce substructure
   with higher frequency in the outskirts of clusters rather than in 
   their centre, and 
c) the asymmetries observed in the ICM would have to be independent 
   of the cD position.

In our model a strong and compact X--ray source
is associated with the dominant galaxy which in turn moves 
about the bottom of the cluster potential well with harmonic oscillations,
producing the observed offset. Whether this compact source is a cooling
flow or an active nucleus is not essential and will be discussed in 
the last section.
Such a model provides a natural explanation of 
the facts summarized above without invoking any dynamical instability.
As a consequence, the fraction of structured clusters is reduced from more 
than $70\%$ (Davis 1994) down to about $30\%$.
\begin{table*}[t]
\caption{{Properties of the cluster sample and of the respective PSPC 
observations.}\label{tab1}}

\centering
\begin{tabular}{c|cccccc}

Abell number	& z	 & ROR		& Exposure (s) 	& Photons$^{(1)}$ 
& $\dot M^{(2)}$ \\ \hline \hline
85	& ~~~~0.0556~~~~& 800250	& 10238		& 16836  & 108.0  \\
133		& 0.0566 & 800319	& 19404		& 9358   & 110.0  \\
262		& 0.0161 & 800254	& 8686		& 9559.  & 9.4	   \\
400		& 0.0232 & 800226	& 23611		& 11856  & 0.0	   \\
478		& 0.0880 & 800193	& 21969		& 12877  & 736.0  \\
496		& 0.0320 & 800024	& 8857		& 13215  & 134.0  \\
539		& 0.0205 & 800255	& 9646		& 7711  & 2.1	   \\
1060		& 0.0114 & 800200	& 15764		& 36310  & 8.0    \\
1651		& 0.0825 & 800353	& 7429		& 6400  & 0.0	   \\
1795		& 0.0616 & 800105	& 36273		& 43724  & 321.0 \\
1991		& 0.0586 & 800518	& 21261		& 6157  & 37.0  \\
2029		& 0.0767 & 800249	& 12542		& 18989  & 431.0 \\
2052		& 0.0350 & 800275	& 6211		& 7813  & 94.0 \\
2063		& 0.0337 & 800184	& 9763		& 9794 & 35.0 \\
2107		& 0.0421 & 800509	& 8274		& 5508  & 7.1  \\
2142		& 0.0899 & 800551	& 6090		& 8881 & 369.0 \\
2589		& 0.0421 & 800526	& 7289		& 6339  & --	\\
2657		& 0.0414 & 800320	& 18904		& 12487 & 44.0 \\
3562		& 0.0490 & 800237	& 20199		& 13503 & 0.0 \\
3921		& 0.0936 & 800378	& 11997		& 7263  & -- \\
4038		& 0.0292 & 800354	& 3353$^{(3)}$	& 4845  & -- \\
4059		& 0.0488 & 800175	& 5439		& 5876  & -- 
\end{tabular}

\raggedright

\footnotesize{
$^{(1)}$ Approximate number of photons (background subtracted) used in the 
large scale centroid determination

$^{(2)}$ Cooling flow mass rate in solar masses per year (from Fabian 1994;
White et al. 1997; Allen \& Fabian 1998). A long dash indicates 
clusters for which no data have been found.

$^{(3)}$ This exposure is lower than the 5 ks limit set for the 
sample, but a sufficiently accurate measurement of the centroid
has been possible.}

\end{table*}

The role of cD galaxies and their motion relative to the
bottom of the cluster potential well is still a matter of debate.
Optical observations, which should be best suited for such an analysis,
suffer from low number statistics in the cluster 
and the results are somewhat contradictory. Studies on individual
clusters (Hill et al. 1992; Sharples et al. 1988; Oegerle \& Hill 1992)
find significant peculiar radial motions of the cD with respect to the
mean recession velocity, and in sample of 
25 cD clusters Zabludoff et al. (1990) 
find that ``a substantial fraction of the cD galaxies have
velocities significantly different from the mean of their parent clusters'',
a result confirmed by Malumuth et al. (1992).
However it is not completely clear whether the presence of substructure in the
velocity field could alter this result (Oegerle \& Hill 1994; Bird 1994).

Numerical simulations of cluster evolution (Malumuth \& Richstone 1984)
predict that cD galaxies are created and spend their existence near the 
centre of clusters and move slowly around them. Moreover, the actual motion 
of the dominant galaxy and the related displacement from the 
bottom of the potential well are also an important clue 
to understand the formation scenario of dominant galaxies 
(see Oegerle \& Hill 1994
and references therein).

By assuming that cD galaxies move about the centre of mass
of their parent clusters we can derive an equation of motion and estimate 
a normalized oscillation amplitude which is independent from the 
parameters of the individual clusters. We can then compare the expected 
distribution of the normalized amplitudes to the  measurements  
from the set of 22 X--ray images of moderate redshift Abell clusters.

\medskip

This paper is organized as follows: section~\ref{data} describes the data
and the reduction procedures, section~\ref{smallsc} and ~\ref{largesc}
describe respectively 
the small and large scale analysis performed on the images, while
section~\ref{offsec} briefly summarize the measured offset
properties. The possible identification of the cD galaxy with the emission
peak is discussed in section~\ref{cdi}; section~\ref{model}
describes the physical oscillatory model 
and the results are summarized and discussed in section~\ref{disc}.

\section{The data: selection criteria and preliminary reduction}
\label{data}

To detect a small but significant offset between the peak 
and the large scale emission of clusters we need a precise determination
of both these positions.

The uncertainty on the determination of the position of a source 
characterized by a profile with a width $\sigma$ 
(irrespective of whether it is intrinsic or due to the instrument
resolution) and a number $N$ of counts is given by:

\be
\Delta(x) \simeq \frac{\sigma}{\sqrt N} \label{spprec}
\ee

\noindent which follows from the standard error propagation
rules. The exact equality holds for a Gaussian spatial 
distribution of photon counts.

Hence, for a point-like source, the best instrument to measure 
the position maximizes the value of the ratio
$\sqrt\epsilon / \sigma_{(\hbox{d})}$,
where $\epsilon$ is the detector efficiency and $\sigma_{(\hbox{d})}$
its intrinsic resolution. However, when the 
angular dimension of the source is intrinsic - as is the case for
the ICM X--ray emission - the width of the photon distribution
is fixed and only a higher 
value of $\epsilon$ can increase the position accuracy. 

Given the above discussion, the cluster observations have been
extracted from the ROSAT/PSPC archive maintained by 
the ROSAT group at the {\it Max Plank Institut f\"ur Extraterrestrische Physik}
in Garching~(D).
We have selected a set of observations pointed on 
Abell clusters with cD galaxy in their centre and a known
value of the radial velocity dispersion. Other constraints were an 
exposure time greater than 5~ks and an angular dimension an order of magnitude
greater than the resolution in the centre of the PSPC field ($\sim 40''$~FWHM).

As of March 1995, we found 22 clusters with ROSAT/PSPC 
pointed observations that met 
the above requirements, listed in table~\ref{tab1}
along with their main properties.
All these observations have $N \gsim 5\times 10^3$ 
counts in the cluster large scale emission. 
With a typical width of $3r_c \simeq 7.5'$ for the clusters as a whole,
equation~\ref{spprec} gives an accuracy of $6''$ in the determination of the
centroid, where $r_c$ is the core radius 
of the modified King function that better fits the cluster profile.
We have used $3 r_c$ as cluster width
to properly take into account the wider wings of the King
with respect to the Gaussian distribution.

The central compact source - which in our model is responsible for
the emission peak - has a mean number of $\gsim 1000$ photons and
a width comparable to the PSPC central resolution of $\sim 25''$ 
($1\,\sigma$).
Equation~\ref{spprec} gives an accuracy $\Delta(x) \lsim 1''$.

The precision on the determination of the peak position could 
be increased using a detector with higher resolution as the ROSAT 
HRI. However, the efficiency of this instrument
is lower than that of the PSPC and hence the accuracy of 
centroid measurement would be decreased. 
Note that to avoid any influence of the boresight errors 
(see section~\ref{cdi}) it is crucial that both the centroid and the peak 
positions are measured on the same image.

\medskip

\begin{figure}[!t]
\psfig{figure=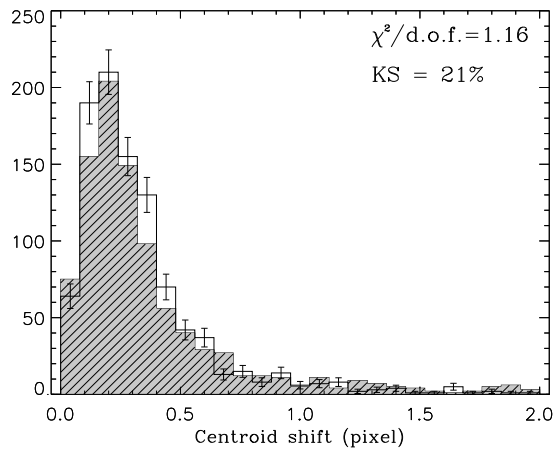,width=8.5cm}
\caption{{Influence of the compact source on the centroid determination.
Histogram of the distance between input and output centroids
are plotted for the simulations with (shaded region) and without
(solid line) the offset compact source. The distributions
are consistent with being two realizations of the same parent.
See the text for the statistical test results.}
\label{sim1ps}}
\end{figure}

\begin{figure}[!t]
\psfig{figure=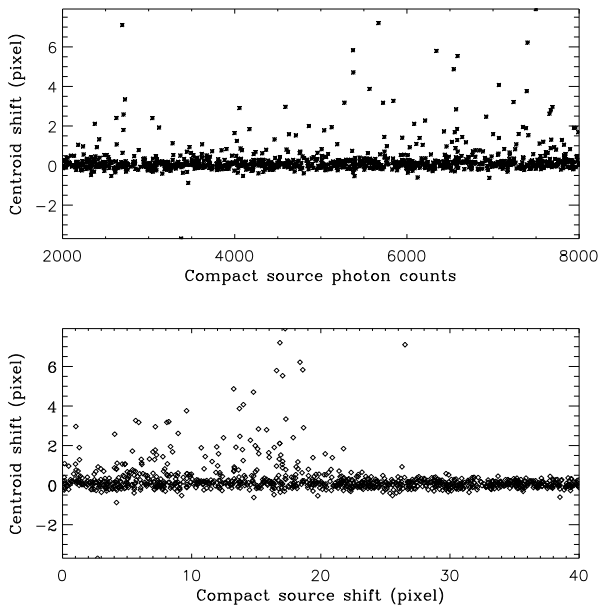,width=8.5cm}
\caption{{Influence of the compact source on the centroid determination.
Measured x-axis errors on the centroid position
versus the photon counts in the offset peak are plotted in the upper
panel.
The vast majority of measurement have a dispersion lower than one pixel,
but a limited number of centroid measurement are shifted towards the 
compact source (positive values in both
panels). No clear dependence of this shift on the power 
of the compact source is seen. 
The lower panel show the same position errors in the measured centroid versus
the offset of the input compact source. In this case a trend with
the compact source centroid is visible. The trend disappears for offsets
larger than 20 pixels, almost the size of the core radius of the
large scale emission.
Note that in both cases the errors in the centroid measurements 
tend to mask the effect of displacement between the centroid and the
emission peak.}
\label{sim2ps}}
\end{figure}

The image reduction has been performed with the complete 
treatment included in the ESAS software (Snowden 1995) which
takes into account all the properties
of the PSPC instrument and of the particular observation
(solar contamination and short and long term enhancements of the
cosmic background). This is because instrumental
effects such as vignetting, detector efficiency, particle background etc.
conspire to build a large scale structure in the images. This structure,
if uncorrected, has its centre coincident with the image centre
and could efficiently mimic an offset with the peak of the true emission.

We have tested the complete reduction process 
searching for large scale residuals in PSPC point-like source
images. We found no evidence for such an effect in the analysis 
of a set of 20 images.

Due to different distance and intrinsic cluster dimensions,
we have implemented specific ESAS routines in order to obtain images
with dynamic pixel size. The final images ($512 \times 512$ pixels each) 
have a pixel size (1 pixel~$= 4 \div 10 ''$)
chosen to maintain a fixed (small) ratio between the whole image width 
and the cluster angular dimensions.

To fully explore any possible bias due to the
reduction process we could simulate a set of cluster images
with instrumental effects included and than compare
the cluster parameters obtained after the reduction
with those input in the simulations.
However, this procedure is extremely time consuming and 
any unknown instrumental effect would remain undetected.

\section{Small scale analysis}
\label{smallsc}

Once a cleaned count-rate image has been obtained, a wavelet transform
algorithm is applied to search for small scale structures
embedded in the cluster emission.
We use the algorithm expressly designed to detect and characterize
small structures overlaid on strongly varying backgrounds
presented and discussed in Lazzati et al. (1998).

The wavelet transform algorithm spans scales from half the PSF on-axis
up to about twice the PSF at the edges of the image, with a spacing of
a factor of two between adjacent scales. This procedure guesses the position
of the emission peak and selects a set of ``point sources'' to be
masked in the subsequent large scale analysis (see section~\ref{largesc}).

The emission peak position is then refined fitting 
a bidimensional Gaussian source model to the image in the 
wavelet space. 
Since the covariance matrix highly underestimates uncertainties in correlated 
data sets, the uncertainties on the peak position are estimated in 
$\chi^2$ space. A $\chi^2$ shift of $4.61$ is used in order to obtain 90\% 
confidence intervals for two independent degrees of freedom
(position in both directions).
Again, this procedure is not strictly correct in the presence of 
correlations between data points. Rosati (1995) performed several tests
to see if any systematic error could arise, confirming the reliability
of this procedure.

Note, however, that the uncertainties on the peak position
are in most cases negligible with respect to those in the large 
scale centroid, which contribute the largest uncertainties to the offset
measurement.

\begin{figure}[!t]
\psfig{figure=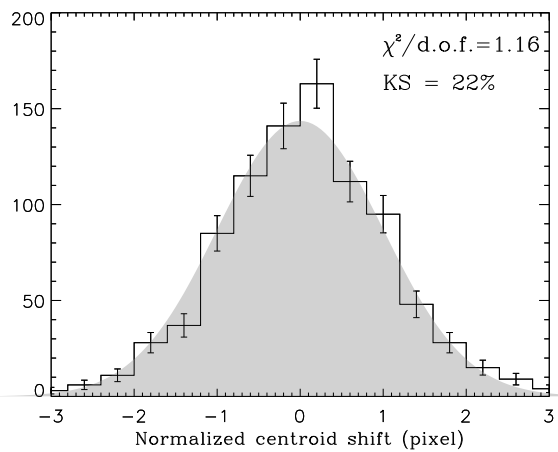,width=8.5cm}
\caption{{
Histogram of the measured errors in recovering the
large scale centroid normalized to the uncertainties
derived from the covariance matrix. The distribution
appears consistent with a Gaussian distribution 
with a normal standard deviation
(shaded area).
}
\label{sim3ps}}
\end{figure}

\section{Large scale analysis}
\label{largesc}

Great care must be used in the determination of the
centroid position since the offset we are looking for is considerably 
smaller than the whole cluster emission.
To obtain the best position for the centroid of the large scale emission,
we fit the cluster emission with an elliptical $\beta$ model:

\be
I_X(r,\theta) \propto {\Big [} 1+ {\Big (} \frac{r}{r_c(\theta)} {\Big )}^2
{\Big ]}^{-3\beta+\frac12},
\label{king}
\ee

\noindent
where $r_c(\theta)$ is the azimuth dependent core-radius. The central 
part, affected by the presence of the com\-pact sour\-ce, was excluded from 
the fit as well as all the point-like sources detected in 
projection over the cluster emission. A bidimensional Gaussian smoothing
of the image was necessary to obtain a more accurate estimate of the 
statistical distribution of points.

To test the reliability of this procedure we have simulated a set of 1000
clusters with properties that span the range of our sample
which are summarized in table~\ref{tab2}.
The core radius of the clusters has been fixed 
to 20 pixels\footnote{This is not a restriction of simulations since
the image size has been chosen to preserve a fixed dimension of the
cluster in pixels.} while all the remaining shape parameters have been
randomly picked up each time from the distributions described in 
table~\ref{tab2}.
To fully reproduce the experimental images, an 
offset Gaussian source ($FWHM=12$ pixels, slightly bigger
than a point-like source)
with a lower number of counts
has been added in the cluster centre and then the
centroid has been measured with the procedure described above.
Analogous simulations have been performed without
the compact offset Gaussian source for comparison.

\begin{table}[!t]
\caption{{Properties of the clusters simulated  to test 
the centroid determination procedure. All distributions are flat except
ellipticity and $\beta$ which have Gaussian probability functions.
}\label{tab2}}
\centering
\begin{tabular}{c|cccccc}
 	& LS$^{(1)}$ &  e $^{(2)}$ & $\beta$ & SS$^{(3)}$ & 
$\Delta(x)$$^{(4)}$ & $\Delta(y)$$^{(4)}$ \\
\hline \hline
average & 15000 & 0.8 & 0.76       & 5000   &  40	  &  40         \\
width   & 10000 & 0.07 & 0.25	   & 3000   &  20          &  20          
\end{tabular}

\raggedright

\footnotesize{
$^{(1)}$ Number of counts in the large scale cluster emission.

$^{(2)}$ Ellipticity defined as the ratio between the minor and major axis.

$^{(3)}$ Number of photon counts in the offset compact source.

$^{(4)}$ Offset of the compact source in the $x$ and $y$ 
directions in pixels with respect to the large scale centroid.
}
\end{table}

Figure~\ref{sim1ps},~\ref{sim2ps} and~\ref{sim3ps} 
show the result of these simulations.
To see whether the presence of the compact offset source
could produce a systematic error in the centroid determination
we have compared the position errors measured from simulations
with and without the compact source.
Figure~\ref{sim1ps} shows the histogram of the distance between input and 
output centroid positions in both cases. The similarity between the
two distributions is confirmed by the Kolmogorov-Smirnov test,
which gives a $\sim 21 \%$ probability of them both
deriving from the same parent distribution. 
Moreover, by fitting the difference of the
two distributions with a zero constant we have obtained a reduced $\chi^2$
value of 1.16.
The absolute values of the distance between the true and the measured
centroid are around a quarter of a pixel, which, given
the resolution of our images, translates in an uncertainty in the
range $1'' - 2.5''$, a value suitable for our
requirements (see section~\ref{data}).

When the offset Gaussian source is present in simulations,
the distribution of the shift between input and output centroids
has a tail that is considerably
higher than what expected in a Gaussian distribution
(the positive values in figure~\ref{sim2ps}).
This effect increases with the shift between the centroid and the peak
position (lower panel of figure~\ref{sim2ps}) up to a sharp
cutoff when the shift becomes larger then the core radius.
We can confidently ignore this effect since 
it tends to reduce the measured offset and not to create it
spuriously.

Finally, we have tested the reliability of the uncertainties 
extracted from the covariance matrix. If these are correct,
the errors measured in the simulations divided by those estimated
in the fitting procedure 
should be distributed as a Gaussian with normal standard
deviation. As shown in figure~\ref{sim3ps} this is what 
actually happens. The observed distribution is consistent
with a Gaussian parent at the 22\% level with a Kolmogorov-Smirnov
test. A fit with a Gaussian gives a reduced $\chi^2$ of 1.17.

\begin{figure}[!t]
\psfig{figure=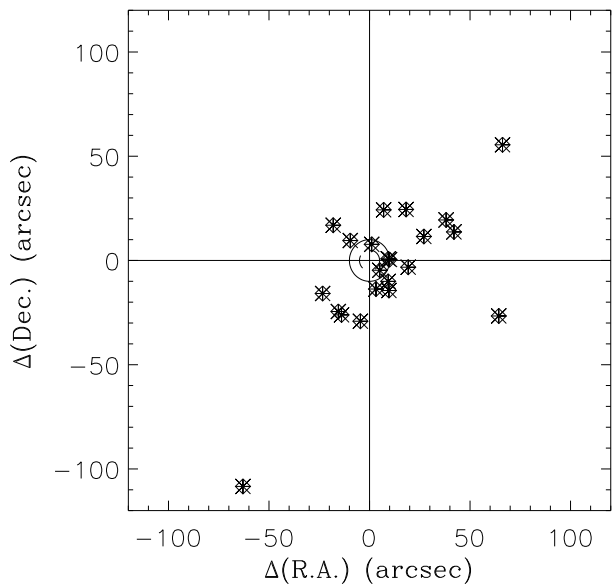,width=9cm}
\caption{{Positions of the large scale centroids with respect 
to the emission peak. The peak 
is located in the origin of coordinates and shifts
in right ascension and declination are given in arcseconds.
The dashed circle encloses the $5''$ confidence region (see text) for the 
centroid position while dotted circle marks the $10''$ region.}
\label{off_ps}}
\end{figure}

\section{Offsets}
\label{offsec}

Applying the procedure described and tested above to the X--ray images
we are able to measure the offsets of the cluster sample along
with their uncertainties. The values of centroid position 
errors estimated in the fitting procedure 
were always $\lsim 5''$ but, to be conservative, we adopt this
value for the whole set of clusters.

\begin{figure}[!t]
\psfig{figure=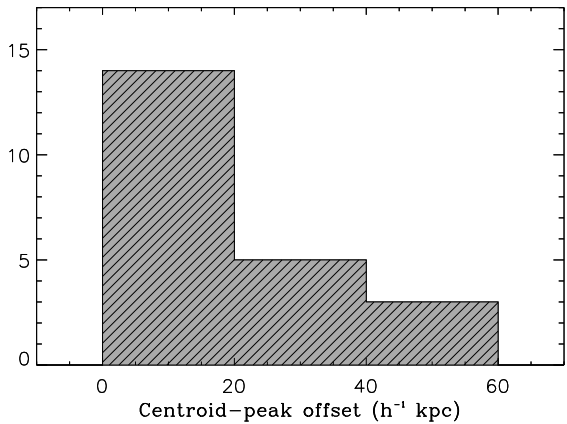,width=9cm}
\caption{{Distribution of the measured offsets in physical units.}
\label{hi_ps}}
\end{figure}

The results are shown in figure~\ref{off_ps}
where the emission peak lie at the center of coordinates
and the symbols mark the position of the centroid. 
The inner circle radius is $5''$ ($1\,\sigma$), while the outer
is $10''$. Only 4 measurements over 22 are in agreement with 
the absence of any offset at the $2\,\sigma$ level.
Figure~\ref{hi_ps} shows the offsets distribution in physical units;
the actual displacements are always lower than $60$ h$^{-1}$ kiloparsec,
a value smaller than the typical core radius of the ICM distribution. 

A final test to see if the characteristics of the detector play a role
in the offset measure can be finally performed on experimental data.
If the measured offsets were
produced by some unknown instrumental effect,
we would expect to find a correlation between the redshift of the cluster
and the offset measured in physical units (kiloparsec).
The linear Pearson correlation coefficient $r$ between
the two dataset is $r=0.25$. The probability that two 
uncorrelated 22 elements datasets give a value of $r > 0.25$
is $P = 40\%$.

\section{X--ray cD counterparts}
\label{cdi}

Once the presence of offsets has been firmly established, 
we can try to correlate both X--ray positions with
the optical position of the dominant cluster galaxy.

Unfortunately, the comparison between optical and X--ray positions
suffer the boresight problem, i.e. an absolute
error in the satellite pointing which turns in a systematic shift
between sources positions and respective optical counterparts.
This error can be corrected when at least 2 bright X--ray
sources in a single image can be correlated with 
bright optical objects (e.g. Guide Star Catalog objects).
If uncorrected, the boresight error is estimated to have a Gaussian 
distribution with standard deviation of $\sim 10''$ both in right
ascension and declination (see the {\it ROSATSRC}
catalog, Zimmermann 1994).

In the A478 image, after the application of a $12''$ boresight correction,
the X--ray peak position is only $3''$ away from the 
cD optical position.

\begin{figure}[!t]
\psfig{figure=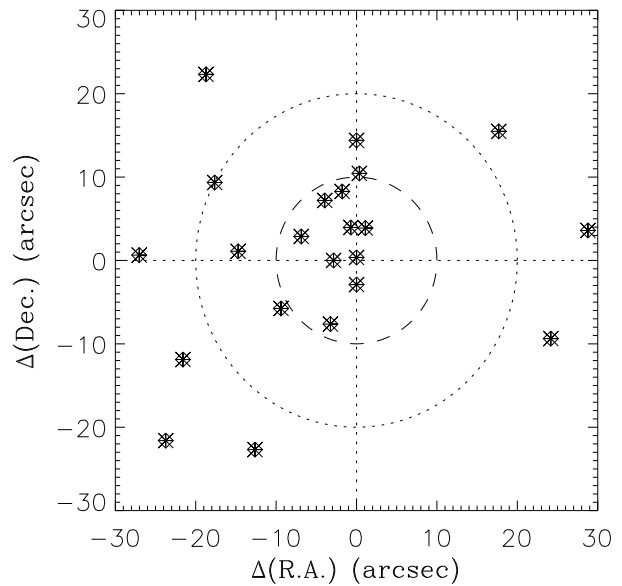,width=9cm}
\caption{{Relative positions of the X--ray peaks with respect to the cD
galaxy. The cD is located in the origin of coordinates and shifts
in right ascension and declination are given in arcseconds.
The dashed circle encloses the $1\sigma$ confidence region for the 
boresight error while
dotted circle marks the $2\sigma$ region.
}\label{off3ps}}
\end{figure}

No X--ray source correlated to a Guide Star Catalog object was found 
in the remaining 21 images and, given the rather large pointing error, 
it has  not been possible to 
establish a firm identification
of the cD galaxy as the object that produces the offset peak
in the large scale X--ray emission of the ICM.
However, we can state that within these uncertainties 
the position of the X--ray peak coincides with the galaxy.
In figure~\ref{off3ps} we show the distribution of X--ray peak
positions around the cD which lies
in the origin of the coordinates. The circles represent $1\sigma$ and
$2\sigma$ confidence regions. As expected, $~40\%$ (9 over 22) of the
X--ray peaks lie inside the $1\sigma$ circle.
The distribution of the Right Ascension residuals in figure~\ref{off3ps}
appears biased towards negative shifts. The tests performed on the
analysis procedure described in section~\ref{largesc} shows that no bias is
present in the X--ray positions and the accuracy of Guide Star Catalogue
is fully adequate; moreover, the absence of the effect in the
declination shifts indicates that the the analysis
procedure is unbiased. The binomial probability of obtaining such a
skewed dataset from a symmetric distribution is $P=3.2\%$.

If the difference between the peak and X--ray
positions were not due to instrumental effects
we would expect a correlation between angular offset and redshift.
The correlation coefficient between these two datasets
is $r = -0.11$.

The possible identification of the cD galaxy with 
the peak of the cluster emission has been discussed in several papers.
Allen et al. (1995) find a coincidence of the two positions in clusters
that host a massive cooling flow in their centre, but do not find
evidence of centroid shift in their sample; more often, a coincidence between
the orientation of the cD isophotes with the ICM emission major
axis is found (Sarazin \& McNamara 1997; Huang \& Sarazin 1998).

Given the identification, we can discuss a model that
predicts the amount of separation between the X--ray peak and the
bottom of the potential well of the ICM, marked by the centroid 
of large scale emission.

\section{The oscillatory model}
\label{model}

An implication of the above discussion is that 
the cD galaxy is not located in the bottom of the 
potential well and hence cannot be at rest but 
must oscillate around the equilibrium position.
This wobbling produces the observed offset.

With a small set of reasonable assumptions and/or approximations on 
cluster structure, we can obtain the equation of motion of the cD 
in the bottom of the potential well.
First, since dominant galaxies are found well inside
the core radius of clusters (section~\ref{offsec}),
we assume that the mass inside the area
of oscillation is uniformly distributed. 

In clusters that host a massive cooling flow in their centres
great deviations from uniformity could be present in the baryon fraction, 
however 
they will have a small effect on the total mass distribution and 
hence do not invalidate our model.
We have therefore:

\bea
m_G\ddot{x} &=& F = -G\frac{M_{int}\, m_G}{x^2} \label{eqmot} \nonumber \\
&=& -\frac43 \pi \, G \, \rho_{tot} \, m_G \, x \label{motieq}
\eea

\noindent
where $m_G$ is the mass of the galaxy, $M_{int}$ the mass contained in a 
sphere of radius equal to the position of the cD and $\rho_{tot}$ the
total mean mass density in the central part of the cluster. 
Solving equation~\ref{motieq} we obtain:

\be
x(t)=B \, \cos(\omega t+\phi) \label{soluz}
\ee

\noindent
where:

\be
\omega=\sqrt{\frac43 \pi \, G \, \rho_{tot}} \label{omega}
\ee

The actual values of $B_j$ and $\phi_j$ for the {\it j-th} cD galaxy
are linked to initial 
speed and position. Since we are concerned with quantities averaged on
an ensemble of clusters, we expect the values of the phases $\phi$ to be
erased.
The value of $B$ can be obtained by assuming the {\it rms} speed of the
galaxy proportional to the radial velocity dispersion of the whole cluster.
This seems to be a reasonable assumption in view of the fact that
the kinematic of the cD, whatever it is, must be related to the 
cluster potential well.

Putting this relation in our model we have:

\bea
\sigma_v^2&=& <v^2>-<v>^2 = <v^2> \nonumber \\
&=& B^2\, \omega^2 <\sin^2(\omega t)> \label{veldisp}
\eea

\noindent
from which:

\be
B^2=\frac{2\sigma_v^2}{\omega^2}=\frac32\,\frac{\sigma_v^2}{\pi \, 
G \, \rho_{tot}}
\label{bvalo}
\ee

We can hence define an adimensional oscillation amplitude $\Gamma$
normalized to the individual cluster properties that should be distributed
as an ensemble of identical harmonic oscillators with the same amplitude but 
with random phases and seen from random orientations.

\be
\Gamma = \sqrt{\frac{2\pi G}{3}} \frac{czd_{\theta}}{H_0} 
\frac{\rho_{tot}^{1/2}}{\sigma_v}
\label{gamm1}
\ee

\be
p(\Gamma) \propto \int_{\Gamma}^L \frac{dl}
{\sqrt{(l^2-\Gamma^2)(L^2-l^2)}}
\label{prob}
\ee

\noindent
where $L$ is the largest oscillation amplitude allowed.

\begin{figure}[!t]
\psfig{figure=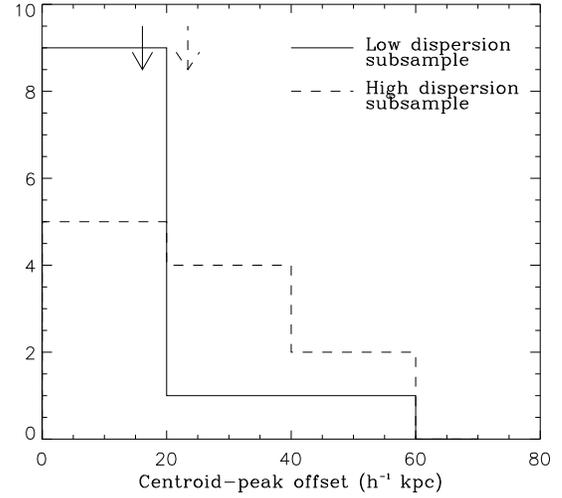,width=8.5cm}
\caption{{Dependence of the physical offset from the radial
velocity dispersion. The offset histogram of the high dispersion subsample
($<\sigma_v> = \, 888 km/s$, dashed line and arrow) is significantly larger 
than that of the low dispersion  ($<\sigma_v> = \, 597 km/s$, 
solid line and arrow) one. The arrows mark the mean value of 
the two distributions.}
\label{twops}}
\end{figure}

To test the likelihood of the assumption that the
cD oscillation velocity is linked to the whole cluster velocity
dispersion, made in equation~\ref{veldisp},
we have divided the sample in two subsamples based on their radial velocity
dispersion. We expect the average offset to
be proportional to the average velocity dispersion
in the two subsamples. As is shown in figure~\ref{twops},
denoting with $_h$ and $_l$ respectively the high and low velocity dispersion 
subsamples we have:

\begin{eqnarray}
<\sigma_v>_h = 888 \, km/s &\qquad& <\Delta>_h = 23.4\, h^{-1} kpc \nonumber \\
<\sigma_v>_l = 597 \, km/s &\qquad& <\Delta>_l = 16.1\, h^{-1} kpc \nonumber 
\end{eqnarray}

\noindent from which:

\be
\frac{<\sigma_v>_h}{<\Delta>_h} = 37.9 \quad 
\frac{<\sigma_v>_l}{<\Delta>_l} = 37.1 \quad 
\ee

The major problem in the application of equation~\ref{gamm1} is the 
determination of the central mean mass density.
In fact, the X--ray emission of the ICM gives a direct information 
of the baryon density while what we need is 
an estimate of the total mean density in the cluster centres.

A crude estimate of the mean central density can be 
inferred applying the virial theorem with the measured cluster
parameters $\sigma_v$ and $r_c$:

\be
\sigma_v^2 \simeq G \frac MR \;\;
\Rightarrow \;\;
\rho_{tot} \sim \frac 1G \frac{\sigma_v^2}{R^2} \sim 
\frac 1G \frac{\sigma_v^2}{r_c^2}
\ee

Inserting this density in equation~\ref{gamm1} we obtain:

\be
\label{gamm2} 
\Gamma = \sqrt\frac{2\pi}{3} \frac{czd_{\theta}}{H_0} \frac 1{r_c} =
\sqrt\frac{2\pi}{3} \frac{d_\theta}{r_{c\,\theta}}
\ee

\noindent
where $r_{c\,\theta}$ is the apparent angular core radius.

\bfi[!t]
\psfig{figure=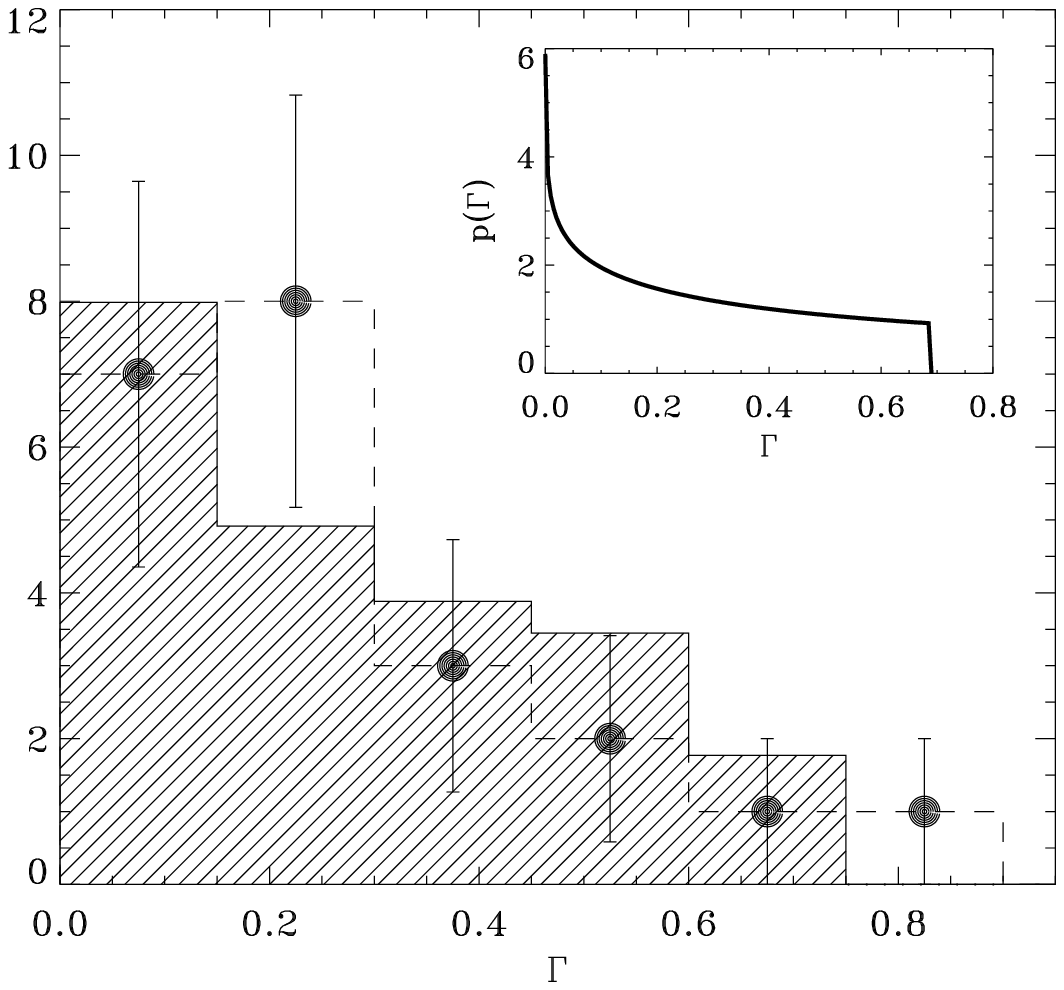,width=8.5cm}
\caption{{Histogram of the values of the $\Gamma$ parameter, as described
in the text. The shaded area represents the expected distribution 
binned with the same width of data. A
Kolmogorov--Smirnov test gives a $\sim 75\%$ probability of success.
The insert shows the theoretical 
distribution (eq.~\protect\ref{prob}) before the binning
process.
Poissonian errors are drawn on experimental points.}
\label{offhist}}
\efi

A further problem that affect the comparison between the harmonic model
and the observations is the intrinsic difficulty
in measuring the core radius of clusters that host a cooling flow
in their centre. This measurement is made highly inaccurate by the need of
masking the central region of the cluster to get rid of the
modification on the cluster central surface brightness due to the
emission of the cool and dense inflowing gas (see, e.g. Allen \& Fabian 1997).
To see which uncertainty this problem could contribute to the 
final parameter $\Gamma$ we have tested the
change of the fitted value of the core-radius increasing
the radius of the masked central region in the image
of A1795.
When the masked region is smaller than the PSF FWHM,
the core radius is highly underestimated. If the mask radius 
is bigger than twice the PSF FWHM - any influence of the central compact 
source being completely erased -
the core radius changes by a factor of  $~30\%$
increasing the mask radius up to 5 times the above value.
This uncertainty propagates with unchanged percentage 
in the determination of the value of $\Gamma$.
Given the smooth shape of the probability distribution (eq.~\ref{prob})
we expect that these uncertainties will affect only the high amplitude
sharp cutoff and the low amplitude cusp of the function, 
leaving the overall shape unmodified (see the insert in figure~\ref{offhist}).

Figure~\ref{offhist} shows the distribution of the measured 
$\Gamma$ in our cluster sample, compared with 
the theoretical distribution obtained rebinning equation~\ref{prob}
(shaded area).
The best value for the largest amplitude $L=0.69$ has been obtained
maximizing the Kolmogorov-Smirnov probability
that the measured distribution is drawn from that theoretically derived.
The maximum probability is $P_{K-S} \simeq 73\%$.
Figure~\ref{offhist} may suggest a slightly bigger value for the 
maximum amplitude $L$, however this would require a higher number of small
offset clusters. This discrepancy can be ascribed to the crudeness of the
model and, in particular, the introduction of circular (or even elliptical)
cD orbits would erase the cusp of the distribution
in the lowest offset region. Such a refinement of the model would
however require a larger dataset to allow any quantitative comparison.

\section{Conclusions and discussion}
\label{disc}

We have analyzed the structure of the X--ray emission of a set
of 22 Abell cD clusters. We find evidence for
a displacement between the centroid of the large
scales and the peak of the emission, the latter being
coincident with the dominant galaxy position.

The presence of centroid shifts in clusters with cD galaxies
and cooling flows is problematic since such offsets are expected to 
be due to subcluster mergers, phenomena that would disrupt
cooling flows (Meiksin 1988; Friaca 1993). 

We have shown that, identifying the peak with the cD
galaxy, an harmonic oscillator model for its motion
in the cluster potential well is in agreement with
present observations and with the predictions 
of the numerical simulations (Malumuth \& Richstone 1984). 
It follows that offset peaks in X--ray images are produced
by this oscillation and should not be thought as signs 
of dynamical instability and/or substructure.

From this interpretation, it follows that the compact emission
in the centre (or near the centre) of clusters can be
produced in principle inside the central galaxy or by a fraction
of the ICM strongly linked to the galaxy itself.
The first hypothesis can be easily discarded since 
observations of cluster central regions with high resolution
instruments show that the compact source
has a small but non-zero extension and hence cannot be due
to an active nucleus (see, e.g., Grebenev et al. 1995, Pierre 
\& Starck 1998).
In the second scenario 
the cool and dense gas of the cooling flow could be 
responsible for the peaked emission seen in X--ray images
and the offset would be caused by the motion of the
cD galaxy. In this case the cooling flow - or at least its inner
part - would have to follow the cD galaxy in its motion.
Whether this is possible or such a motion would disrupt 
the frail flow stability has to be investigated in 
more detail. From the observational point of view,
Pierre \& Starck (1998) have analyzed a sample of high resolution
images of clusters, finding that in the inner cores of
massive cooling flow clusters the X--ray emission 
shows peculiar features and strong isophotal twisting
(see their figure 6).

As a further consequence of this interpretation, the fraction of 
structured clusters is greatly reduced. In fact, in his analysis of
a flux limited sample of X--ray clusters, Davis (1994) found a 
residual emission near the centre of clusters 
(but not coincident with them) 
after subtraction of the extended component with the
elliptic isophotal fitting technique. The interpretation of this
feature as a sign of substructure led to the conclusion
that up to $70\%$ of the X--ray clusters were structured.
If their results are reinterpreted on the light of the conclusions
of this work, we find that the above fraction falls to
less than $30\%$.

\medskip

The hardest problem we had to deal with in this work
is represented by pointing errors in X--ray images. The boresight 
uncertainties do not allow us to place tight constraints 
on any significant displacement between the peaks of the X--ray and 
optical emission and, even if figure~\ref{off3ps} implies the
positions are in  agreement, any improvement will require
better astrometric accuracy through new observations.
Note, however, that the comparison with numerical
simulations and the calculation of the parameter $\Gamma$
described in section~\ref{model} have been performed
only using relative positions from the X--ray images and hence do not
suffer of any systematic pointing error.

The new generation of X--ray telescopes (JET-X, AXAF, XMM, 
WFXT, ...),
thanks to their higher spatial resolution and sensitivity, 
will presumably give us the possibility of better testing the model.
Moreover, it is desirable that 
new observations will allow to enlarge the set of cluster
observations to be compared to the predicted 
harmonic amplitude.

\begin{acknowledgements}
We thank the referee, D.A. White, for pointing out a theoretical
argument which led to an improvement in this work and for
many useful comments and suggestions.
We thank S. Campana and S. Covino for fruitful discussions
and for carefully reading the manuscript.
\end{acknowledgements}

\end{document}